\documentclass[
 a4paper,
 reprint,
 showpacs,
 preprintnumbers,
 amsmath,
 amssymb,
 amsfonts,
 aps,
 prb,
 showkeys,
 superscriptaddress,
 floatfix,
 titlepage
]{revtex4-1}

\usepackage{newtxtext}
\usepackage[varg,cmbraces,cmintegrals]{newtxmath}
\usepackage{bm}
\usepackage{graphicx}
\usepackage{dcolumn}
\usepackage{ifpdf}

\usepackage{color}
\definecolor{darkred}{rgb}{0.6,0,0}
\definecolor{darkblue}{rgb}{0,0,.6}
\definecolor{darkgreen}{rgb}{0,0.5,0}

\ifpdf
\usepackage{epstopdf}
\usepackage[pdftex,unicode,pdfstartview={FitH},pdfborder={0 0 0}]{hyperref}
\usepackage{hypcap}
\else
\usepackage[hypertex]{hyperref}
\fi
\hypersetup{
  bookmarksnumbered = true,
  colorlinks = true, linkcolor = darkblue,
  citecolor = darkblue, filecolor = darkblue,
  menucolor = darkblue, urlcolor = darkblue
}

\newcolumntype{C}{>{$\displaystyle}c<{$}}
\newcolumntype{R}{>{$\displaystyle}r<{$}}

\begin{document}

\title{Valley-selective energy transfer between quantum dots in  atomically thin semiconductors}

\author{Anvar~S.~Baimuratov}
  \email{anvar.baimuratov@lmu.de}
  \affiliation{Fakult\"at f\"ur Physik, Munich Quantum Center, and Center for NanoScience (CeNS),  Ludwig-Maximilians-Universit\"at M\"unchen, Geschwister-Scholl-Platz 1, D-80539 M\"unchen, Germany}
\author{Alexander~H\"ogele}
  \affiliation{Fakult\"at f\"ur Physik, Munich Quantum Center, and Center for NanoScience (CeNS),  Ludwig-Maximilians-Universit\"at M\"unchen, Geschwister-Scholl-Platz 1, D-80539 M\"unchen, Germany}
  \affiliation{Munich Center for Quantum Science and Technology (MCQST),
  Schellingtra\ss{}e 4, 80799 M\"unchen, Germany}
 
\begin{abstract}
In monolayers of transition metal dichalcogenides the nonlocal nature of the effective dielectric screening leads to large binding energies of excitons. Additional lateral confinement gives rise to exciton localization in quantum dots. By assuming parabolic confinement for both the electron and the hole, we derive model wave functions for the relative and the center-of-mass motions of electron--hole pairs, and investigate theoretically resonant energy transfer among excitons localized in two neighboring quantum dots. We quantify the probability of energy transfer for a direct-gap transition by assuming that the interaction between two quantum dots is described by a Coulomb potential, which allows us to include all relevant multipole terms of the interaction. We demonstrate the structural control of the valley-selective energy transfer between quantum dots.
\end{abstract}


\maketitle

\section{Introduction}
The unique properties of two-dimensional (2D) materials provide versatile opportunities for nanomaterial physics \cite{Geim2007}. Within this realm, monolayers of transition metal dichalcogenides (TMD) represent 2D crystalline semiconductors with unique spin and valley physics for opto-valleytronic applications \cite{Xu2014,Schaibley2016,mak2016}. Coulomb electron--hole attraction and the nonlocal nature of the effective dielectric screening lead to a large binding energy of excitons, which dominate both light absorption and emission \cite{Chernikov2014,He2014,Ye2014,wang_colloquium_2018}. The combination of exceptional brightness and spin--valley coupling opens up novel opportunities for tunable quantum light emitters for quantum information processing and sensing \cite{Chakraborty2019} realized on the basis of excitons confined in TMD-based systems.

There are various approaches to realize exciton confinement in TMD monolayers. Impurities, vacancies, or strain in monolayer TMD crystals, as well as local modulations of the immediate environment, modify the energy gap of 2D materials \cite{Rhodes2019}. By providing additional lateral confinement, local disorder is known to confine excitons in a relatively small area of TMD monolayers and give rise to quantum dot (QD) excitons.
Spectral signatures of quantum dot exciton localization with bright and stable single-photon emission were observed from unintentional defects in monolayer tungsten diselenide \cite{Srivastava2015a,He2015,Koperski2015,Chakraborty2015,Tonndorf2015}. Subsequently, strain engineering has proven as a viable deterministic approach to obtain spatially and spectrally isolated quantum emitters in monolayer and bilayer TMDs \cite{Kumar2015,Branny2016,Branny2017,Palacios-Berraquero2017}, and controlled positioning has been achieved by irradiating monolayer crystals with a sub-nm focused helium ion beam \cite{Klein2019}. Alternatively, QDs have been realized by electrostatic confinement \cite{Zhang2017,Wang2018,Pisoni2018}, or by creating lateral TMD heterostructures forming a potential well \cite{Huang2014}. In vertical TMD heterostructures, moir\'e superlattices give rise to periodic QD arrays hosting localized excitons \cite{Seyler2019,Tran2019}.

By preserving strong spin--valley coupling, TMD QDs inherit optovalleytronic properties from their 2D host system \cite{Lu2019}, as the intervalley coupling is weak due to the vanishing amplitude of the electron wave function at the QD boundary and hence valley hybridization is quenched by the much stronger spin--valley coupling \cite{Liu2014}. As in conventional QDs, the oscillator strength and radiative lifetime of confined excitons are strongly size-dependent, which results in oscillator strength enhancement and ultrafast radiative annihilation of excitons, varying from a few tens of femtoseconds to a few picoseconds \cite{Fouladi2018}. In the presence of an external magnetic field bound states in TMD QDs can be considered as quantum bits for potential applications in quantum technologies \cite{Korm2014,Pearce2017,Dias2016}. 


In this work, we study nonradiative resonance energy transfer between two adjacent QDs in TMD monolayers \cite{Agranovich2009}. Building on theories initially developed for molecular systems by F\"orster in the framework of dipole--dipole interaction \cite{Forster1948} and generalized by Dexter for quadrupole and exchange interactions \cite{Dexter1953}, we derive the theory of nonradiative resonance energy transfer for atomically thin QDs hosted by 2D crystals. For conventional QDs with sizes in order of tens of nm, the multipole nature of Coulomb interactions and energy transfer through dipole-forbidden states must be taken into account \cite{Kruchinin2008}. For QDs hosted by 2D systems considered here, we account for multipole terms of the transfer process and analyze the effect of the donor--acceptor system geometry on transfer efficiency. 

The manuscript is organized as follows. In the Sec.~\ref{sec_qd} we derive the energy spectrum and wave functions of excitons in TMD QDs with parabolic confinement. Section~~\ref{sec_fret} contains the calculations of the Coulomb matrix elements and the nonradiative resonant energy transfer between two QDs coupled by a Coulomb potential. Concluding remarks are given in Sec.~\ref{sec_conc}.


\section{Model for confined excitons}\label{sec_qd}
We start our analysis from delocalized excitons in TMD monolayer, by focusing on excitons formed by states at the bottom of the conduction band and the topmost valence band at $\mathbf{K}_\pm$ points of the first Brillouin zone.
Using the two-band effective mass model the wave function of the exciton can be written as a factorization of the relative motion of charge carriers and their center-of-mass motion \cite{Courtade2017,Glazov2015}:
\begin{equation}\label{free_exc}
  \tilde{\mathcal F}^{(\alpha)}_{N \mathbf{Q}}(\mathbf{r}_e, \mathbf{r}_h) =
  \sigma^{-1/2} \exp{(i \mathbf{Q R})}\tilde\psi_N (\bm{\rho}) u_\alpha (\mathbf{r}_e) v_{-\alpha} (\mathbf{r}_h),
\end{equation}
where $\alpha = \pm 1$ is the valley index,
$\mathbf{r}_{e (h)}$ is the radius vector of the electron (hole),
$\sigma$ is the normalization area,
$\mathbf{R} = (m_e \mathbf{r}_e + m_h \mathbf{r}_h) / M$ is the center-of-mass vector,
$m_{e(h)}$ is the effective mass of the electron (hole),
$M = m_e + m_h$,
$\mathbf{Q}$ is the wave vector of the center-of-mass motion,
$\tilde\psi_N (\bm{\rho})$ is the wave function of the relative motion with coordinate $\bm{\rho} = \mathbf{r}_e - \mathbf{r}_h$, and
$u_\alpha (\mathbf{r}_e)$ and $v_\alpha (\mathbf{r}_h)$ are the Bloch functions of the electron and hole in valley $\alpha$.

We solve the Schr\"odinger equation for the relative motion of states with circular symmetry, namely $S$-states with zero angular momentum,
\begin{equation}
  \left[
    -\frac{\hbar^2}{2 \mu} \Delta + V (\rho)
  \right]
  \tilde\psi_N (\rho)
  =
  \tilde\epsilon_N \tilde\psi_N (\rho),
\end{equation}
where $\rho = |\bm{\rho}|$, $\mu = m_e m_h / M$ is the reduced mass,
$\tilde\epsilon_N$ is the eigenenergy of the $S$-state with the principal quantum number $N$.
The nonlocally screened electron--hole interaction is described by the Rytova--Keldysh potential \cite{Keldysh1979,Rytova1967,Cudazzo2011}
\begin{equation}\label{Coulomb}
  V(\rho) =
  -\frac{\pi e^2}{2 \varepsilon \rho_0}
  \left[
    H_0 \left( \frac{\rho}{\rho_0} \right) 
    - Y_0 \left( \frac{\rho}{\rho_0} \right) 
  \right],
\end{equation}
where $e$ is the electron charge, $\rho_0$ is the screening length,
$\varepsilon$ is the effective dielectric constant, and
$H_0 (x)$ and $Y_0 (x)$ are Struve and Neumann functions.

\begin{figure}[t]
  \centering
  \includegraphics[scale=1]{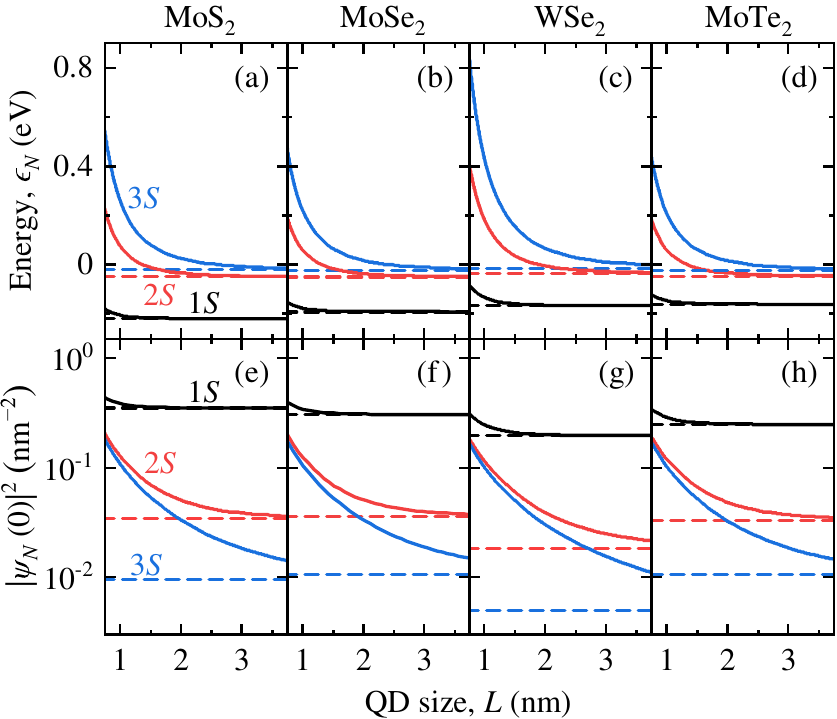}
  \caption{
    QD size effects (solid lines) on the energies $\epsilon_N$ of $1S$ state $(N=1)$, $2S$ state $(N=2)$, and $3S$ state $(N=3)$ for (a)--(d) MoS$_2$, MoSe$_2$, WSe$_2$, and MoTe$_2$ monolayers with material parameters from Table~\ref{tab01} and $\varepsilon = 4.5$.
    (e)--(h) Same for the overlaps of the electron and hole wave functions at the same spatial position $|\psi_N (0)|^2$.
    Dashed lines show the limit of delocalized excitons with the energies $\tilde\epsilon_N$ and overlaps $|\tilde\psi_N (0)|^2$.
  }
  \label{fig01}
\end{figure}

Without considering the details of lateral confinement we focus on TMD QDs with in-plane localization of charge carriers and make the approximation of a harmonic confinement.
If the coordinates of the relative motion, $\bm{\rho}$, and the center-of-mass motion, $\mathbf{R}$, of the confined exciton are not separated, one can use a variational procedure without the separation of coordinates to find the wave function \cite{Semina2007}.
Here we assume for simplicity that both the electron and hole are confined by parabolic potentials of the form
\begin{equation}\label{parabola}
  U (\mathbf{r}_e, \mathbf{r}_h) =
  \frac{\omega^2 }{2} \left(m_e \mathbf{r}_e^2 + m_h \mathbf{r}_h^2\right) =
  \frac{\omega^2 }{2} \left(\mu \rho^2 + M R^2\right),
\end{equation}
which are characterized by the confinement frequency $\omega$.
With this potential we separate the coordinates of the relative motion and the center-of-mass motion \cite{Que1992}. 
Therefore, the energies and wave functions of the excitons in QDs are written as:
\begin{align}
  \label{energy_weak}
    \mathcal E_{N,n l} &=
    E_g + E_{n l} + \epsilon_N,
  \\  
  \label{exc_wf_weak}
    \mathcal F^{(\alpha)}_{N,n l} (\mathbf{r}_e, \mathbf{r}_h) &=
    \Psi_{n l} (\mathbf{R}) \psi_N (\bm{\rho})
    u_\alpha (\mathbf{r}_e) v_{-\alpha} (\mathbf{r}_h),
\end{align}
respectively.
The total energy of the exciton confined in the QD takes discrete values and is dependent on the band gap of the TMD monolayer, $E_g$, and on $E_{n l}$ and $\epsilon_N$, which are the energies of the confined center-of-mass and relative motions.

We solve the Schr\"odinger equation for the center-of-mass motion
\begin{equation}
  \left[
    -\frac{\hbar^2}{2 M} \Delta + \frac{M \omega^2 R^2}{2}
  \right]
  \Psi_{n l} (\mathbf{R}) 
  =
  E_{nl} \Psi_{n l} (\mathbf{R})
\end{equation}
in polar coordinates $\mathbf{R} \equiv (R,\theta)$.
The exact eigenenergies and eigenstates are those of the 2D harmonic oscillator \cite{Landau2013}, namely
\begin{equation}\label{energy_exc_qd}
    E_{nl} =
    (2n + |l| + 1) \hbar \omega,
\end{equation}
and
\begin{subequations}
  \begin{align}
  \label{trans_wf}
    \Psi_{nl} (\mathbf{R}) &=
    L^{-1} A_{nl} \mathcal{R}_{nl}(R/L) \Phi_l (\theta),
  \\
    A_{nl} &=
    \sqrt{2\cdot n!/(|l| + n)!},
  \\
    \mathcal{R}_{nl}(x) &=
    x^{|l|} e^{-x^2/2} \mathcal L^{|l|}_n (x^2),
  \\
    \Phi_l (\theta) &=
    e^{i l \theta}/\sqrt{2 \pi},
  \end{align}
\end{subequations}
where $n = 0, 1, 2, ...$ and $l = 0, \pm 1, \pm 2, ...$ are the principal and angular momentum quantum numbers, respectively, $L = \sqrt{\hbar/(M \omega)}$ is the QD size, and  $\mathcal L^{|l|}_n (x)$ are the associated Laguerre polynomials.

\begin{table}[tb]
  \caption{
    The effective masses of electrons and holes (in units of the free electron mass) and 2D screening lengths are taken from Refs.~\onlinecite{Korm2015,Han2018}.
  }
  \begin{ruledtabular}
  \begin{tabular}{cCCCC}
    Parameter & \mathrm{MoS_2} & \mathrm{MoSe_2} & \mathrm{WSe_2} &\mathrm{MoTe_2} \\
    \hline
    $m_e$                      & 0.45    & 0.53  & 0.34  & 0.57 \\
    $m_h$                      & 0.54    & 0.6   & 0.36  & 0.64 \\
    $\rho_0$ (\AA)             & 6.67    & 10    & 8.2   & 14.4  \\
  \end{tabular}
  \end{ruledtabular}
  \label{tab01}
\end{table}

The radial part of the relative motion with zero angular momentum is determined not only by the nonlocally screened potential from Eq.~\eqref{Coulomb} as in the case of free excitons, but also by the parabolic potential $\mu \omega^2\rho^2/2$. The Schr\"odinger equation for the relative motion is written as
\begin{equation}
  \left[
    -\frac{\hbar^2}{2 \mu} \Delta
    + \frac{\mu \omega^2  \rho^2}{2}
    + V (\rho)
  \right] \psi_N (\rho)
  =
  \epsilon_N \psi_N (\rho).
\end{equation}
To find an approximate solution of this equation one can use the 2D hydrogen-like wave functions with the Bohr radius as variational parameter \cite{Chernikov2014}.
Using the material parameters from Table~\ref{tab01}, we solve this equation numerically for the first three exciton $S$-states in four specific TMDs, namely MoS$_2$, MoSe$_2$, WSe$_2$, and MoTe$_2$.

The energies $\epsilon_N$ and wave function overlaps of the electron and hole at the same spatial position $|\psi_N (0)|^2$ are shown in Fig.~\ref{fig01} as functions of the QD size $L$.
In the top panel of Fig.~\ref{fig01} we observe for all materials the same trends for the energies $\epsilon_N$, they decrease with the size of the QD.
The $1S$ state $(N=1)$ is less dependent on the QD size as it is the most localized state.
The wave function overlaps $|\psi_1(0)|^2$ are shown in Figs.~\ref{fig01}(e)--\ref{fig01}(h).
Due to the confinement effect these overlaps are larger for smaller QDs and decrease monotonically with size.
QDs in WSe$_2$ exhibit larger size effect than in MoS$_2$, MoSe$_2$, and MoTe$_2$ due to smaller reduced mass.
For large QDs all three states converge to those of delocalized excitons shown by the dashed lines in Fig.~\ref{fig01}.

\section{Resonant energy transfer}\label{sec_fret}

In the following, we calculate the nonradiative resonant energy transfer between two 2D QDs coupled by a Coulomb potential.
Due to the finiteness of QD dimensions in the $xy$-plane, the point dipole model developed by F\"orster leads to large errors when the distance between the QDs is of the order of their sizes.
Therefore, the multipole nature of Coulomb interaction must be taken into account as in the case of conventional 3D QDs \cite{Kruchinin2008}.
For conventional colloidal QD and molecular systems the orientations of QDs and molecules are random, whereas in layered systems the positions of the QDs are fixed and the localized excitons with lowest energies have an in-plane circular polarization with sign reversal for $\mathbf{K}_+$ and $\mathbf{K}_-$.
This in-plane arrangement of the dipole moments leads to characteristic valley-selective orientation effects, which are considered in detail below.

\begin{figure}[tb]
  \centering
  \includegraphics[scale=1]{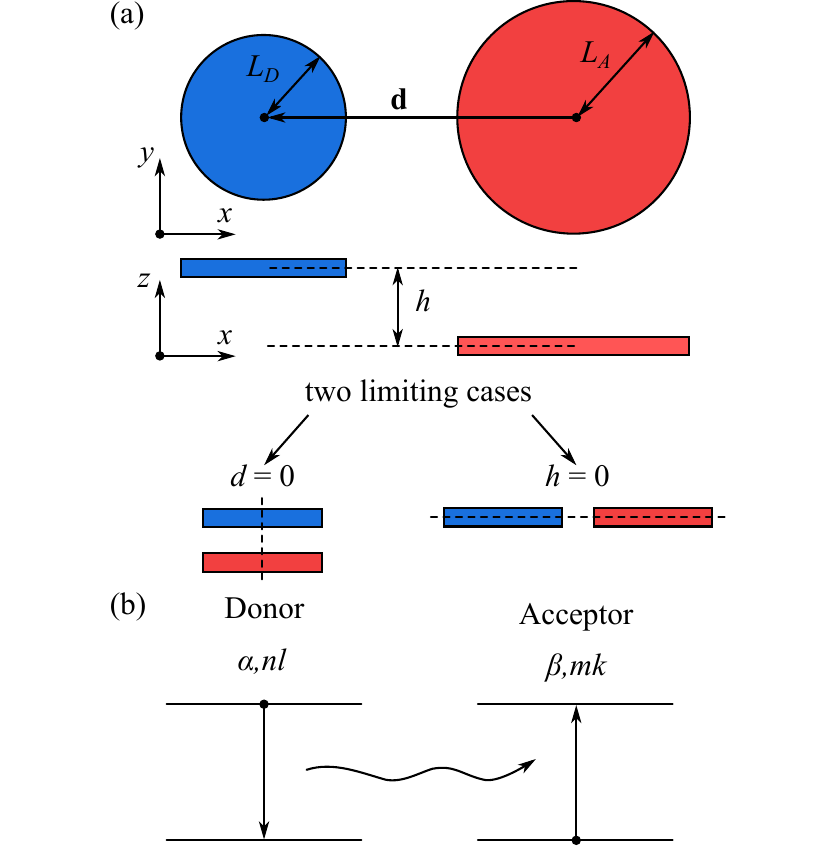}
  \caption{
    (a)~Schematic illustration of two TMD QDs, donor and acceptor. The geometries of the two limiting cases: the centers of two QDs lie on the $z$-axis (left) or two QDs lie in one $xy$-plane.
    (b)~Nonradiative resonant energy transfer between donor and acceptor.
  }
  \label{fig02}
\end{figure}

Let us consider two 2D QDs, namely donor and acceptor with sizes $L_D$ and $L_A$, located in the same or in two different layers of the same TMD material.
We assume the centers of QDs to be separated by a 2D-vector $\mathbf{d}$ in the $xy$-plane as shown in Fig.~\ref{fig02}(a).
In the $z$-direction they are separated by a distance $h$ as illustrated in the $xz$-plane projection.
In our analysis, it is useful to distinguish two limiting cases.
The first case corresponds to the situation when the QDs lie in different layers on top of each other and $d = 0$. 
The second case is realized when the QDs are located in one monolayer and $h = 0$.
These two limiting cases allows us to analyze the orientation effects, \textit{e.g.}, for $h = 0$ the system is a truly 2D-object, in which the QDs and their dipole moments are in one plane.
For $d = 0$ the problem is quasi-3D, because in principle all dimensions matter.

Without loss of generality we assume that only the $1S$ states $(N = 1)$ contribute to the energy transfer between the donor and acceptor.
Then, energy transfer is related to the annihilation of an exciton $\mathcal F^{(\alpha)}_{1, n l}$ in the donor and the creation of an exciton $\mathcal F^{(\beta)}_{1, m k}$ in the acceptor [see Fig.~\ref{fig02}(b)], where $\alpha$ and $n l$ $(\beta$ and $m k)$ are the valley index and two quantum numbers of the exciton in the donor (acceptor).
Here we consider the range of distances between the donor and acceptor, $s = (d^2 + h^2)^{1/2}$, much smaller that the de Broglie wavelength of the annihilated exciton and neglect effects of exchange and radiation transfer \cite{Poddubny2016,Samosvat2016}, but take into the account the multipole nature of the Coulomb interaction \cite{Kruchinin2008}.
With these assumptions, the energy transfer depends on the matrix elements of the Coulomb potential
\begin{equation}\label{matrix_simple}
  C_{nl,mk}^{\alpha\beta} =
  \frac{e^2}{\varepsilon}
  \iint \mathrm{d} \mathbf{r}_1 \mathrm{d} \mathbf{r}_2
  \frac{\mathcal F^{(\beta)*}_{1, m k} (\mathbf{r}_2, \mathbf{r}_2) \mathcal F^{(\alpha)}_{1, n l} (\mathbf{r}_1, \mathbf{r}_1)}
  {\sqrt{(\mathbf{d} + \mathbf{r}_1 - \mathbf{r}_2)^2 + h^2}},
\end{equation}
where $\mathbf{r}_1$ and $\mathbf{r}_2$ are the 2D center-of-mass vectors originating at the centers of the donor and acceptor.
By using the Fourier expansion we find the matrix elements 
\begin{equation}\label{q_matrix}
  C_{nl,mk}^{\alpha\beta}  =
  \frac{e^2}{\varepsilon} \frac{|\psi_1(0)|^2}{2 \pi}
  \int \mathrm{d} \mathbf{q}
  \frac{e^{i \mathbf{q d}-qh}}{q} F_A^* (\mathbf{q}) F_D (\mathbf{q}),
\end{equation}
where
\begin{equation}
  F_D (\mathbf{q}) =
  \int \mathrm{d} \mathbf{r} e^{i \mathbf{q r}} \Psi_{n l} (\mathbf{r}) u_\alpha (\mathbf{r}) v_\alpha (\mathbf{r}).
\end{equation}
If we use the long wave approximation $qa \ll 1$ with the lattice constant of TMD $a$ and express the integration as a sum of integrals over elementary cells, we simplify 
\begin{equation}\label{ft_env}
  F_D (\mathbf{q}) =  
  i^{2n+|l|+1} \sqrt{2 \pi} L_D  A_{nl} \mathcal{R}_{n l} (q L_D)\mathbf{q r}_{v c}^{(\alpha)},
\end{equation}
where the interband matrix element of the coordinate operator is
\begin{equation}
  \mathbf{r}_{v c}^{(\alpha)} =
  \frac{1}{\Omega} \int_{\Omega} \mathrm{d} \mathbf{r} v_{-\alpha} (\mathbf{r}) \mathbf{r} u_\alpha (\mathbf{r})
\end{equation}
and $\Omega$ is the area of the unit cell.
The corresponding expression for $F_A (\mathbf{q})$ can be found by replacements $D\rightarrow A$, $\alpha \rightarrow \beta$, and $nl \rightarrow mk$ in Eq.~\eqref{ft_env}.

By carrying out the integration over the angular variable of $\mathbf{q}$ in Eq.~\eqref{q_matrix} we obtain 
\begin{equation}\label{matrix_rv}
  C_{nl,mk}^{\alpha\beta}  =
  \frac{e^2}{\varepsilon} 
  \left[
    \mathcal{I}_1
    \mathbf{r}_{c v}^{(\beta)} \mathbf{r}_{v c}^{(\alpha)}
    - \mathcal{I}_2
    \left( \mathbf{n r}_{c v}^{(\beta)} \right)
    \left( \mathbf{n r}_{v c}^{(\alpha)} \right)
  \right]
\end{equation}
with two integrals $\mathcal{I}_1$ and $\mathcal{I}_2$ given by
\begin{multline}\label{integrals}
  \mathcal{I}_\eta  =
  2\pi i^{2(n-m)+|l|-|k|} A_{nl} A_{mk} L_A L_D  |\psi_1(0)|^2 d^{-2}
  \\
  \times
  \int_0^\infty \mathrm{d} q (q d)^\eta e^{-q h} J_\eta (q d) 
  \mathcal{R}_{m k} (q L_A)
  \mathcal{R}_{n l} (q L_D),
\end{multline}
where $J_\eta(x)$ is the Bessel function of the first kind, $\eta = 1, 2$, and $\mathbf{n}$ is the unit vector co-directional with $\mathbf{d}$.

Within the two-band approximation of the band structure in TMD monolayer, the interband matrix elements of the dipole moment operator in the donor and acceptor are written in Cartesian coordinates as
\begin{align}
  e \mathbf{r}_{c v}^{(\alpha)}   =
  e \mathbf{r}_{v c}^{(-\alpha)} &=
  \frac{\mathcal{D}}{\sqrt 2} e^{i \alpha \varphi} (1 ,i \alpha), \\
  e \mathbf{r}_{c v}^{(\beta)}    =
  e \mathbf{r}_{v c}^{(-\beta)}  &=
  \frac{\mathcal{D}}{\sqrt 2} e^{i \beta \vartheta} (1 ,i \beta),
\end{align}
where $\mathcal{D} = e a t / E_g$ and $t$ is the nearest-neighbor hopping integral \cite{Xiao2012,Gartstein2015}, and angles $\varphi$ and $\vartheta$ determine the crystal coordination axes of the donor and acceptor.
Using these formulas for dipole moments and choosing $\mathbf{n} = \mathbf{e}_x$, we find
\begin{equation}\label{matrix_fin}
  C_{nl,mk}^{\alpha\beta}  =
  \frac{p\mathcal{D}^2}{\varepsilon}
  \left(
    \delta_{\alpha \beta} \mathcal{I}_1 - \mathcal{I}_2/2
  \right),
\end{equation}
where $\delta_{\alpha\beta}$ is the Kronecker symbol and $p = e^{i (\beta\vartheta-\alpha\varphi)}$ is the phase factor, which is determined by the alignment of the crystal coordination axes.
Further, for the sake of simplicity, we assume $\varphi = \vartheta = 0$ and $p = 1$. According to the result, the \textit{intravalley} matrix element $(\alpha = \beta)$ corresponds to the annihilation and creation of excitons in the same valley and is proportional to the difference $\mathcal{I}_1 - \mathcal{I}_2 /2$.
The \textit{intervalley} matrix element $(\alpha \neq \beta)$ is proportional to $- \mathcal{I}_2/2$, because the dipole moments of the excitons are orthogonal to each other and the first scalar product in Eq.~\eqref{matrix_rv} is zero.

Finally, using the Fermi's golden rule we obtain the rate of the resonant energy transfer from the donor state $\alpha,nl$ to all final states of acceptor $\beta,mk$ as:
\begin{equation}\label{gen_trans}
  \gamma_{nl}^{(\alpha)} =
  \frac{2}{\hbar^2} 
  \sum_{\beta mk} \left|C_{nl,mk}^{\alpha\beta}\right|^2 \frac{\Gamma}{\Gamma^2 + \Delta^2},
\end{equation}
where $\Gamma$ is the sum of the total dephasing rates of interband transitions in the donor and acceptor, and $\hbar\Delta = E_{nl}-E_{mk}$ is the energy detuning between the exciton levels in the donor and acceptor involved in the energy transfer process.
Notably, if the magnitudes of the Coulomb matrix elements in Eq.~\eqref{gen_trans} are much larger than $\hbar\Delta$ and $\hbar\Gamma$, the formation of the entangled states in QDs must be considered \cite{Kruchinin2015}.

\begin{figure}[t]
  \centering
  \includegraphics[scale=1]{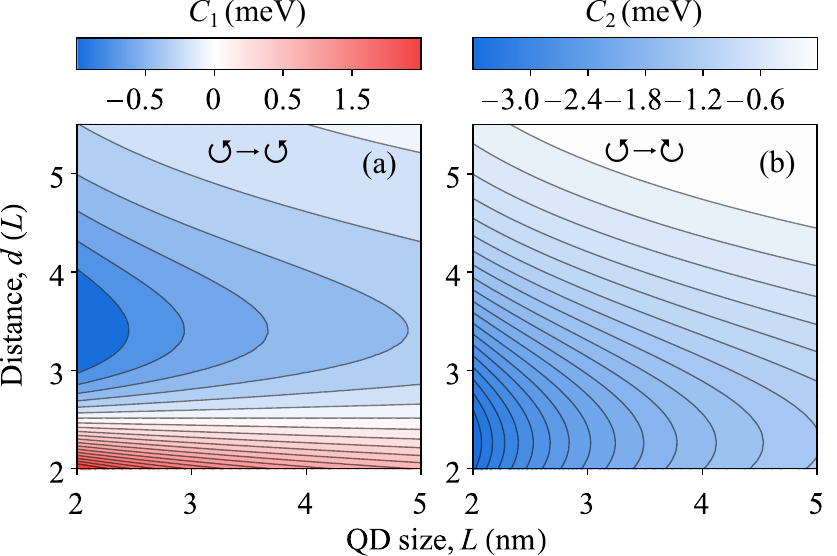}
  \caption{
    Contour map of (a) the intravalley and (b) the intervalley Coulomb matrix elements for QDs in MoS$_2$ monolayer as a function of their size $L$ and distance $d$.
    The material parameters of monolayer MoS$_2$ are $a = 3.193$~\AA, $t = 1.56$~eV, and $E_g = 1.85$~eV \cite{Tongay2014,Xiao2012}.
  }
  \label{fig03}
\end{figure}

Assuming the simplified case when the energy transfer occurs for equal QDs, $L_D = L_A = L$, from the state $nl = 00$ in valley $\alpha$ to the states with $mk = 00$ in valleys $\beta = \pm\alpha$, we find
\begin{equation}\label{transfer}
  \gamma_{00}^{(\alpha)} \approx
  \gamma =
  \frac{2}{\hbar^2\Gamma}
  \left(
    \left|C_{00,00}^{\alpha\alpha}\right|^2 +
    \left|C_{00,00}^{\alpha,-\alpha}\right|^2
  \right),
\end{equation}
where the rate is dependent on the intravalley and the intervalley matrix elements.
To quantify these matrix elements for $h > 0$ and $d > 0$ we evaluate the integrals in Eq.~\eqref{integrals} numerically.

Before proceeding with the evaluation of the integrals, it is instructive to introduce the dipole--dipole approximation (DDA) of the Coulomb interaction developed by F\"orster \cite{Forster1948}. This approximation of point dipoles is obtained by increasing the distance between the donor and acceptor $s$ to infinity, in particular $d \rightarrow \infty$ for $h = 0$ or $h \rightarrow \infty$ for $d = 0$. Then we find the DDA limit for the matrix element from Eq.~\eqref{matrix_fin} and the energy transfer rate from Eq.~\eqref{transfer}
\begin{align}
  \label{DDA_matrix}
  C^\mathrm{(dd)} &= 
    \lim_{s \rightarrow \infty} C
    \propto s^{-3}, 
  \\
  \label{DDA_trans}
  \gamma^\mathrm{(dd)} &= 
    \lim_{s \rightarrow \infty} \gamma
    \propto s^{-6},
\end{align}
where the matrix element indexes were omitted for simplicity.

As illustrative examples of our theory, we consider two limiting cases shown in Fig.~\ref{fig02}(a), as they allow us to analyze the matrix elements and rates, and find their asymptotics.
For $h = 0$, when both QDs are located in the same monolayer, we find exact expressions
\begin{align}
  \label{eq_c1}
  C_1 &\equiv C_{00,00}^{\alpha\alpha}|_{h=0} =
  \frac{\sqrt \pi C_0}{4} 
  \frac{(1 - 2\xi) I_0 (\xi) + 2\xi I_1 (\xi)}
  {\exp{(\xi})},
  \\
  \label{eq_c2}
  C_2 &\equiv C_{00,00}^{\alpha,-\alpha}|_{h=0} =
  \frac{\sqrt \pi C_0}{4} 
  \frac{(1 + 2\xi) I_1 (\xi)- 2\xi I_0 (\xi)}
  {\exp{(\xi})},
\end{align}
where $C_0 = 2\pi \mathcal{D}^2 |\psi_1(0)|^2/(\varepsilon L)$, $\xi = d^2/(8 L^2)$, and $I_0 (\xi)$ and $I_1 (\xi)$ are the modified Bessel functions of the first kind.
In Fig.~\ref{fig03} we show the matrix elements for MoS$_2$ monolayer and $\varphi = 0$ starting from the close-contact distances between QDs, $d = 2 L$.
For the shown range of distances the intravalley matrix element $C_1$ first starts with a positive value and decreases with distance $d$.
After crossing the zero at $d \approx 2.51 L$ it reaches a minimum $C_1 \approx -0.03 C_0$ at $d \approx 3.41 L$ and further increases monotonically.
On the other hand, the intervalley matrix element $C_2$ starts from a negative value, decreases monotonically with distance, and after reaching a minimum $C_2 \approx -0.15 C_0$ at $d \approx 2.26 L$ it increases monotonically.

The analysis shows that for $\xi \rightarrow \infty$, which is the DDA in Eq.~\eqref{DDA_matrix} for $h = 0$, we obtain
\begin{equation}
  3C_1^\mathrm{(dd)} = C_2^\mathrm{(dd)} = -3 C_0 (L/d)^3.
\end{equation}
For small distances in the monolayer limit ($\xi \ll 1$) exchange effects must be taken into account.

\begin{figure}[tb]
  \centering
  \includegraphics[scale=1]{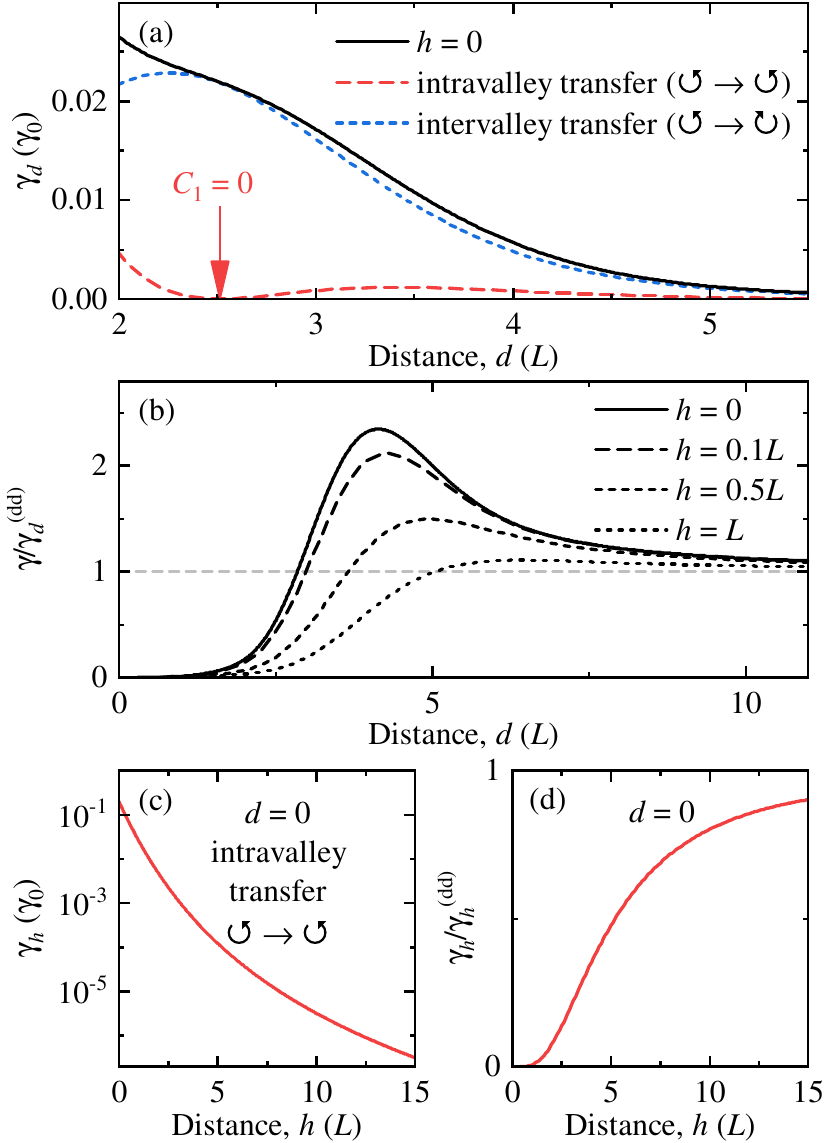}
  \caption{
    (a)~Energy transfer rate between two QDs in one monolayer, $h = 0$ (solid line).
    The dashed and short-dashed lines show the intravalley and intervalley contributions.
    The red arrow shows the distance $d$, which corresponds to the pure intervalley transfer.
    (b)~Ratio between energy transfer rates $\gamma/\gamma_{d}^\mathrm{(dd)}$ for fixed distances $h = 0,0.1L,0.5L,L$.
    (c)~Energy transfer rate between two QDs for $d = 0$.
    (d)~Ratio between energy transfer rates $\gamma/\gamma_{h}^\mathrm{(dd)}$ for a fixed distance $d = 0$.
  }
  \label{fig04}
\end{figure}

By substitution of Eqs.~\eqref{eq_c1} and \eqref{eq_c2} into Eq.~\eqref{transfer} we find the rate of the energy transfer for $h = 0$ as
\begin{multline}\label{gamma_d}
  \gamma_d =
  \frac{\pi \gamma_0}{16}
  e^{-2 \xi}
  \Big[
  (1 - 4\xi ) \left(I_0^2(\xi) + I_1^2(\xi)\right) \\
  + 8\xi^2 \left(I_0(\xi) - I_1(\xi)\right)^2
  \Big],
\end{multline}
where $\gamma_0 = 2 C_0^2 / (\hbar^2 \Gamma)$.
By using the material parameters of monolayer MoS$_2$ and assuming $L = 3$~nm, $\varepsilon = 4.5$ and $\hbar\Gamma = 5$~meV \cite{Klein2019}, we estimate the absolute values $C_0 = 17$~meV and $\gamma_0 = 176$~ps$^{-1}$.
Again for $\xi \rightarrow \infty$ we obtain the DDA in Eq.~\eqref{DDA_trans} $\gamma_{d}^\mathrm{(dd)} = 10 \gamma_0 (L/d)^6$.
This result clearly shows that for this limit, the system can be considered as two interacting point dipoles in the F\"orster model. 
We plot the rate of the energy transfer $\gamma_d$ in Fig.~\ref{fig04}(a) and show explicitly the intravalley and intervalley contributions to the transfer.
Evidently, the intervalley transfer is larger than the intravalley, particularly for large distances it is nearly one order of magnitude larger.
It should be noted, that for the distance $d \approx 2.51 L$ only the intervalley transfer occurs, this position is marked by the red arrow in Fig.~\ref{fig04}(a).

To compare the exact result for the energy transfer in Eq.~\eqref{gamma_d} with the DDA we calculate the ratio between the energy transfer rates $\gamma/\gamma_{d}^\mathrm{(dd)}$ in Fig.~\ref{fig04}(b).
The gray dashed line corresponds to the DDA asymptotic.
For the monolayer limit with $h = 0$ we use Eq.~\eqref{gamma_d}, whereas for QDs separated in the $z$-direction with distances $h = 0.1L, 0.5L, L$ we substitute Eqs.~\eqref{matrix_fin} and \eqref{integrals} into Eq.~\eqref{transfer}.
Evidently, for small distances between the QDs the multipole contribution becomes sizable and the distance dependence deviates from the DDA.
It is a result of the finite sizes of the QDs in the  $xy$-plane.
With increasing distance $d$ the ratios $\gamma/\gamma_{d}^\mathrm{(dd)}$ become larger than unity, and also have maxima, \textit{e.g.} for the monolayer limit we observe the maximum $\gamma_d \approx 2.35 \gamma_{d}^\mathrm{(dd)}$ at $d \approx 4.14 L$.
With increasing separation in the $z$-direction, $h$, we observe a decrease in the energy transfer and a shift of the ratio maximum to larger distances $d$.
Summarizing the above, the monolayer limit exhibits the largest energy transfer rate, that is up to 2.35 times larger than the energy transfer for the DDA. 

Another limiting case with $d = 0$ is shown in Fig.~\ref{fig02}(a) and corresponds to the geometry, where the centers of the two QDs lie on the $z$-axis.
For this symmetry only the intravalley energy transfer occurs, and its matrix element is given by
\begin{equation}
  C_{00,00}^{\alpha\alpha}|_{d=0} =
  \frac{C_0}{4}
  \left[
    \sqrt{\pi} e^{\zeta^2} \left(1 + 2\zeta^2\right) \mathrm{Erfc}(\zeta)-2\zeta
  \right],
\end{equation}
where $\zeta = h/(2L)$ and $\mathrm{Erfc} (\zeta)$ is the complementary error function.
Using Eq.~\eqref{transfer} we calculate the transfer rate as
\begin{equation}
  \gamma_h =
  \frac{\gamma_0}{16}
  \left[
    \sqrt{\pi} e^{\zeta^2} \left(1 + 2\zeta^2\right)  \mathrm{Erfc}(\zeta)-2\zeta
  \right]^2.
\end{equation}
From the asymptotic behavior of these functions for $\zeta \rightarrow \infty$, which corresponds the DDA in Eqs.~\eqref{DDA_matrix} and \eqref{DDA_trans} for $d = 0$, we obtain $2 C_0 (L/h)^3$ for the intravalley matrix element and $\gamma_{h}^\mathrm{(dd)} = 4 \gamma_0 (L/h)^6$ for the energy transfer rate. In Fig.~\ref{fig04}(c) we show that with increasing distance $h$ the transfer rate $\gamma_h$ decreases subexponentially. For all distances this ratio is smaller than unity and the DDA overestimates the energy transfer. This holds generally if all multipole contributions are taken into account, which is necessary when the distance $s$ is in the order of the QD sizes and the energy transfer deviates from the F\"orster model.

\section{Summary and conclusions}\label{sec_conc}
In summary, we developed a simple theory for excitons confined in QDs of atomically thin TMD semiconductors. Using the approximation of harmonic confinement we derived the energy spectrum and the wave functions of QD excitons. By calculating the intravalley and intervalley Coulomb matrix elements and taking into account not only dipolar contributions but also multipole corrections, we determined resonant energy transfer rates between two adjacent QDs. We derived exact expressions for two simplified cases of possible donor--acceptor geometries, and found that for small distances all multipoles must be taken into account. At large distances, the energy transfer was found to converge asymptotically towards the F\"orster DDA. The largest energy transfer rate was found in the monolayer limit, where the major contribution stems from the intervalley matrix element of the Coulomb interaction and the intravalley matrix element is small. Moreover, the intravalley matrix element can be drastically suppressed by geometry, rendering the energy transfer valley selective. This aspect is important when considering collective light-matter effects such as superradiance for QD ensembles, and might prove useful for spin-valley selective transfer of quantum information.

\begin{acknowledgments}
The authors thank M.\,M. Glazov, M.\,A. Semina, and I.\,V. Martynenko for fruitful discussions. A.\,S.\,B. has received funding from the European Union's Framework Programme for Research and Innovation Horizon 2020 (2014--2020) under the Marie Sk{\l}odowska--Curie Grant Agreement No.~754388 and from LMU Munich's Institutional Strategy LMUexcellent within the framework of the German Excellence Initiative (No. ZUK22). A.\,H. acknowledges funding by the European Research Council (ERC) under the Grant Agreement No.~772195, and the Deutsche Forschungsgemeinschaft (DFG, German Research Foundation) under Germany's Excellence Strategy EXC--2111--390814868.
\end{acknowledgments}

\bibliography{lit}
\end{document}